# Anisotropic non-split zero-energy vortex bound states in a conventional superconductor

Howon Kim[1,*], Yuki Nagai[2,3], Levente Rózsa[1,§], Dominik Schreyer[1], and Roland Wiesendanger[1,*]

[1]Department of Physics, University of Hamburg, D-20355 Hamburg, Germany

[2]CCSE, Japan Atomic Energy Agency, Kashiwa, Chiba 277-0871, Japan

[3]Mathematical Science Team, RIKEN Center for Advanced Intelligence Project (AIP), Chuo, Tokyo 103-0027, Japan

Corresponding authors(*):  hkim@physnet.uni-hamburg.de, wiesendanger@physnet.uni-hamburg.de

[§]Current address: Department of Physics, University of Konstanz, D-78457 Konstanz, Germany

**Abstract:**

**Vortices in topological superconductors are predicted to host Majorana bound states (MBSs) as exotic quasiparticles. In recent experiments, the spatially non-split zero-energy vortex bound state in topological superconductors has been regarded as an essential spectroscopic signature for the observation of MBSs. Here, we report the observation of anisotropic non-split zero-energy vortex bound states in a conventional elemental superconductor with a topologically trivial band structure using scanning tunneling microscopy and spectroscopy. The experimental results, corroborated by quasi-classical theoretical calculations, indicate that the non-split states directly reflect the quasiparticle trajectories governed by the surface electronic structure. Our study implies that non-split zero-energy states are not a conclusive signature of MBSs in vortex cores, stimulating a revision of the current understanding of such states.**

# I. INTRODUCTION

Majorana bound states (MBS) in condensed matter systems have attracted considerable interest as a key element for fault-tolerant topological quantum computation [1–3]. They have been theoretically predicted to exist in topological superconductors, which may be realized by low-dimensional magnetic nanostructures, semiconductor nanowires, or topological insulators proximitized to s-wave superconductors [4–8]. However, due to experimental difficulties in the fabrication of these sophisticated nano-scale hybrid systems, significant efforts have been directed towards studies of intrinsic topological superconducting bulk materials, being a promising platform for hosting MBSs in the cores of vortices which form in the presence of an external magnetic field [9]. Recently, several materials have emerged as potential topological superconductors possessing spin-polarized topological surface states, including iron-based superconductors [10–13] and transition-metal dichalcogenides [14]. MBSs are expected to be observed at zero energy (i.e., the Fermi level) in the center of the vortex cores. In particular, the observation of spatially non-split zero-energy states in the local density of states (LDOS) emanating from the vortex core using scanning tunneling microscopy and spectroscopy (STM/STS) has been regarded as an essential experimental fingerprint for MBSs in these materials [10–12,14].

The identification of MBSs based exclusively on spectroscopic signatures is made challenging by the simultaneous presence of topologically trivial bound states. These include Yu-Shiba-Rusinov (YSR) states in magnet-superconductor hybrid systems [15–17], Andreev bound states in semiconductor nanowire – superconductor hybrid systems [18], and Caroli-de Gennes-Matricon bound states (CBS) in vortex cores [19]; the latter were also observed in topologically trivial superconductors such as $NbSe_2$ [20,21] and on the (110) surface of bulk Nb [22]. CBSs may theoretically be distinguished from MBSs through their spatial distribution around the vortex: they appear at finite but non-zero energy in the core, and their energy increases away from the core, leading to a splitting in the spectra with increasing distance. Based on these properties, the simultaneous observation of the anisotropic non-split and the split zero-energy states has been interpreted as a signature for the coexistence of the MBSs and CBSs at a single magnetic vortex [14,23].

Despite these recent advances, many open questions still remain concerning the observation of MBSs in vortex cores. Although angle-resolved photoemission spectroscopy (ARPES) measurements identified the non-trivial topology of several iron-based superconductors in reciprocal space, zero-energy bound states have only been observed in a fraction of vortices by STS in real space. It has also been realized that distinguishing the spectral features of CBSs from those of MBSs in the vortex core may be impossible if the energy separation of the subgap bound states is below the experimental resolution limit, but the difference between spatially splitting and non-splitting states as a criterion for identifying MBSs has not been challenged so far. However, the splitting of the CBSs with distance is based on the assumption of an isotropic electronic structure for the superconductor, while the materials studied in Refs.14 and 23 have been demonstrated to possess an anisotropic Fermi surface. Calculations of the quasiparticle LDOS at vortices based on the quasi-classical approximation have revealed the crucial role of the anisotropy of the superconducting order parameter and the Fermi surface for its spatial distribution [24,25].

Here, we present an experimental proof for the presence of anisotropic non-split zero-energy bound states in the vortex cores of the conventional and topologically trivial superconductor lanthanum based on STM and STS investigations. The observed spatial distribution of the zero-energy states is reproduced by quasi-classical theory calculations taking into account the anisotropic Fermi surface formed by the quasi-two-dimensional surface states of La(0001) [26,27], thereby identifying them as topologically trivial CBSs. Accordingly, non-split zero-energy states cannot longer be considered as evidence for the existence of MBSs.

## II. RESULTS AND DISCUSSION

### A. STM/S characterization of the La(0001) surface and visualization of the vortex bound states

Bulk-like $\alpha$-lanthanum films with thickness larger than the superconducting coherence length ($\xi_{lit} = 36.3$ nm [28]) were epitaxially grown on a (0001)-oriented rhenium single crystal substrate [See the Appendix for more details]. Superconducting La-coated PtIr tips were prepared to improve the energy resolution in tunneling spectroscopic measurements beyond the Fermi-Dirac limit [27,29,30]. All STM/STS measurements were carried out at T=1.65 K, which is well below the superconducting transition temperature of both the La films and the La-coated tip, $T_c \sim 5$-6 K [31].

We first characterized the atomic and electronic structure of the atomically flat surface of the La(0001) film in zero magnetic field. In Fig. 1(a), a local tunneling conductance (dI/dV) spectrum obtained on the clean La(0001) surface shows a clear resonance peak at +110 meV reflecting the band edge of the surface states originating from the $d_{z^2}$-like atomic orbitals of La [26,27], together with a superconducting gap feature at the Fermi energy. This peak feature is considerably different from the V-shaped dip in the vicinity of the Fermi level, which has been interpreted as a signature of a Dirac cone in the dispersion of topological surface states [10,14]. The calculation of the symmetry-based indicators (see Methods and Refs.32 and 33) of the band structure of La does not predict the presence of topologically protected surface states, either. According to our calculations within the framework of density-functional theory (DFT), the observed surface states of La(0001) crossing the Fermi level are almost spin degenerate, with a negligible Rashba splitting on the order of a few meVs. In contrast to this, the theoretical description of MBSs at vortex cores requires the formation of an effective p-wave pairing, facilitated by a single spin-polarized band crossing the Fermi level. The inset of Fig. 1(a) shows an atomic-resolution STM image with the hexagonal arrangement of the surface La atoms. To resolve the surface electronic structure in momentum space, we measured energy-dependent differential tunneling conductance (dI/dV) maps on La(0001). In Fig. 1(b), we present a dI/dV map obtained at V = +2.0 mV showing quasiparticle interference patterns resulting from surface electron scattering at residual atomic defects on the La(0001) surface. The corresponding Fourier transformation (FT) of the dI/dV map in Fig. 1(c) clearly shows a hexagonally shaped pattern which can be attributed to the anisotropic Fermi surface contour of the quasi-2D surface electronic band [27]. $\mathbf{Q}_1$ and $\mathbf{Q}_2$ denote the scattering vectors along the high-symmetry crystallographic directions where the scattering of surface quasiparticles occurs towards the vertex and the flat part of the hexagon, respectively. In Ref. [27], it was demonstrated using FT analysis at various energies in agreement with *ab initio* calculations that the hexagonal Fermi contour originates from the intra-band scattering of the quasi-2D surface states, leading to the large spatial extension of YSR bound states around magnetic impurities on the La(0001) surface.

Next, we characterized the superconductivity of the La sample and the STM tip, and verified the presence of vortex bound states by applying a magnetic field B perpendicular to the La(0001) surface. The field-free tunneling spectrum (black, solid) at the bottom of Fig. 1(d) reflects the characteristics of a superconductor-superconductor (S-S) tunnel junction showing strong coherence peaks at $E_{coh} = \pm|\Delta_S + \Delta_T| = \pm 1.6$ mV, where $E_{coh}$ is the energy of the coherence peaks and $\Delta_{S\,(T)}$ is the superconducting gap of the La surface (tip). Note that $\Delta_T/e$ (*e* is the electron charge) corresponds to the Fermi energy on the La(0001) surface, i.e., zero energy. By simulating the spectrum (orange, dotted) considering the superconducting state of the La-coated tip, we deduce values of $\Delta_S = 0.95$ meV and $\Delta_T = 0.65$ meV. They have been used for obtaining the deconvoluted surface DOS without the contribution of the tip DOS as shown in Fig. 1(e) [34]. By applying a magnetic field of B=0.1 T perpendicular to the surface, we created vortices penetrating the La film. As superconductivity is quenched within the vortex cores, resulting in a reduced intensity of the coherence peaks, we could visualize the spatial distribution of the vortices as depressions around the vortex centers by measuring the dI/dV map at $E_{coh}$ as shown in Fig. 1(f). At the center of individual vortices, we observe pronounced resonance peaks in the tunneling spectrum (red curve in Fig. 1D) at $V_S$=+$\Delta_T/e$, i.e., equivalent to zero energy (red curve in Fig. 1(e)), which are absent from the spectrum obtained away from the vortices (blue curves in Fig. 1(d) and 1(e)). The pronounced resonance peaks in the vortex cores can be attributed to zero-energy vortex bound states or CBS, which, in principle, are due to the coherent superposition of the Andreev reflected electron and hole from the inhomogeneous superconducting pair potential around the vortex. The spatial distribution of the CBS can be visualized in the dI/dV map at $V_S = +\Delta_T/e$ showing an anisotropic star shape extending more than ~100 nm along the six directions equivalent to the $\mathbf{Q}_2$ direction (Fig. 1(g)). A star-shaped LDOS modulation around vortices has been observed previously for the conventional multi-gap superconductor $NbSe_2$ [20,21].

## B. Spatial distribution of the vortex bound states

In order to get more insight into the energy-dependent spatial distribution of the anisotropic vortex bound states, we present the spatial evolution of the LDOS around an individual vortex based on the tunneling spectroscopic measurements in Fig. 2(a)-2(d) (see also Supplementary Movie 1 [34]). At zero energy (E=0 or $V_S$=+$\Delta_{tip}/e$, Fig. 2(a)), the vortex has an anisotropic star shape with six rays extending from the center of the vortex towards the directions of the six flat parts of the Fermi contour ($\mathbf{Q}_2$) as shown in Fig. 1(c). As the energy is increased, each ray is split into two and they move away from the center while keeping the direction of the extension. Since the separation between two split rays becomes larger and larger at higher energy, this leads to the expansion of the vortex and simultaneously to the appearance of a depression of the LDOS at the vortex center as shown in Figs. 2(b)-2(d).

Figures 2(e)-2(g) show the spectral evolutions of the LDOS along the three radial lines with $\theta$ = 0°, 15° and 30° passing through the center of the vortex as the origin, respectively. All spectroscopic maps show highly symmetric evolutions with respect to the vortex center. At $\theta$ = 0°, where the radial line is parallel to the $\mathbf{Q}_1$ direction, the zero-energy peak of the vortex bound state at the center is split into two dispersive inner and outer branches, i.e. a linear and a curved X-shaped splitting is observed away from the center. Remarkably, the splitting of the linear branch becomes smaller by increasing $\theta$ (Fig. 2(f)). Finally, as shown in Fig. 2(g), the inner lines merge into a single line, namely, a spatially non-split zero-energy state along the radial direction with $\theta$ = 30°, parallel to $\mathbf{Q}_2$ pointing towards the flat parts of the Fermi contour. On the other hand, the outer branch exhibits nearly the same curved X-shaped splitting for all values of $\theta$. In Fig. 2(h), the angle-dependent LDOS plot along a circular line (white dotted line in Fig. 2(a), 30 nm away from the vortex center) clearly shows that the spatially non-split zero-energy states appear at every radial line for $\theta$=30°+N·60° due to the hexagonal symmetry (with N an integer), while it shows energy splitting elsewhere, $\theta$≠30°+N·60°.

Although a similar anisotropic extension of spatially non-split zero-energy bound states has been interpreted as an essential signature of MBSs in 2M-WS$_2$ and Bi$_2$Te$_3$/FeTe$_{0.55}$Se$_{0.45}$, this explanation can be ruled out here due to the topologically trivial electronic structure of La. The pronounced zero-energy vortex bound state can be identified as a CBS. The star-shaped LDOS modulation around the vortex has been similarly observed in a conventional multi-gap superconductor, NbSe$_2$ [20,21], and the modulation of the LDOS has been demonstrated to be anisotropic in vortices on the (110) surface of bulk Nb [22]. Possible explanations of the anisotropic extension include vortex-vortex interactions, or the anisotropy of either the Fermi surface or of the superconducting gap, regardless of topological properties. The influence of vortex-vortex interactions in the present system appears to be negligible, since the extension of the bound states is hardly influenced by the inhomogeneities in the vortex lattice, as shown in Fig. 1(g) and 1(f). The field-free tunneling spectrum in Fig. 1(d) shows a nearly ideal single-component BCS-type superconducting gap, excluding the importance of the gap anisotropy.

## C. Role of surface electronic structure onto the vortex bound states

To elucidate the major role of the Fermi surface for the spatial distribution of the vortex bound states, we calculated the LDOS around a single vortex based on a quasi-classical Green's function method within the Kramer-Pesch approximation (KPA), which is suitable for the low-energy quasiparticle excitations within the vortex core [35]. The difference between the energy levels of CBSs in the quantum limit is $\Delta_S/k_F\xi$, which is approximately 10 μeV based on the measured $\Delta_S$=0.95 meV (see Fig. 1(d)), $k_F \approx 2\pi \cdot 0.5$ nm$^{-1}$ (based on Fig. 1(c)) and $\xi$=36.3 nm for bulk La. Since this is well below the experimental resolution, CBSs appear indistinguishable from zero-energy resonances, and the quasi-classical approximation is justified. Energy-resolved QPI measurements together with DFT calculations of the surface states indicated that the magnitude of the Fermi velocity $v_F$, and consequently the coherence length $\xi$, is similar along the $\mathbf{Q}_1$ and $\mathbf{Q}_2$ directions [27], but the quasiparticle trajectories are focused along the latter due to the shape of the Fermi contour.

Figures 3(a)-3(d) show the calculated results of the quasiparticle DOS maps around an isolated superconducting vortex by considering the realistic Fermi contour of the La(0001) surface (the inset of Fig. 3(a)). The nearly-hexagonal Fermi contour of the La(0001) surface has been extracted from first principles and can be compared with the quasiparticle interference pattern at the Fermi energy (see Fig. 1(c) and Ref. [27]). Remarkably, the six-fold star-shaped extension observed at zero energy (Fig. 3(a)) splits into twelve pairwise parallel rays at higher energies (Figs. 3(b)-3(d) and see also Supplementary Movie 2 [34]), reflecting the trajectory of scattered quasiparticles with a higher impact parameter measured from the vortex center. Since the strongly anisotropic Fermi surface selects preferential directions for the quasiparticle velocities, the trajectories primarily run parallel to the $\mathbf{Q}_2$ and symmetrically equivalent directions at all energies. We found an enhancement of the quasiparticle DOS at the positions where the extension lines are interconnected with each other. Furthermore, Figs. 3(e)-3(g) show the calculated LDOS maps as a function of energy and distance from the vortex center along radial lines for $\theta$=0, 15 and 30° angles measured from the $\mathbf{Q}_1$ direction, respectively (see also Supplementary Movie 3 [34]). As the crossing point of quasiparticle trajectories with an angle of 60° between them is moved away from the vortex center with increasing energy, a cross-shaped pattern is formed in the spectral evolution for $\theta$=0° with increasing CBS energies away from the center, as reflected by the high-intensity hexagons in Figs. 3(b)-3(d). The branches of the cross approach each other for $\theta$=15° and merge for $\theta$=30°, resulting in a non-split zero-energy bound state. All of the above features of the simulated spectra are in excellent agreement with the experimental observations in Fig. 2. Within the quasi-classical description, the non-split zero-energy state is a reflection of the delocalized quasiparticle trajectories at zero-energy due to the heavily weighted Fermi velocity along the direction where the Fermi contour is flat [27,36]. On the other hand, for an isotropic Fermi contour, the spatial distribution of the vortex bound states is isotropic (or circular) [34]. In addition to the anisotropic distribution of the CBS, the shape of the Fermi surface plays an important role for the long-

ranged extension of the CBS due to the effective reduction of the dimensionality compared to the case of the isotropic Fermi surface [27,34]. Since we found nearly identical spectroscopic signatures for the topologically trivial elemental superconductor La as reported before for other types of superconductors with an anisotropic Fermi surface, this clearly indicates that a non-split zero-energy bound state in a superconducting vortex is not a characteristic signature of a MBS, but rather an effect of the anisotropic Fermi surface on the spatial distribution of the CBS.

## III. SUMMARY

In summary, we have studied the spatially- and energy-resolved LDOS of vortex bound states in a conventional and topologically trivial superconducting La(0001) sample using high-resolution STM/STS. We have shown that the spatial distribution of the vortex bound states reflect the quasiparticle trajectory around the vortex via the anisotropic Fermi surface of La(0001). Anisotropic non-split zero-energy vortex bound states coexisting with split branches at finite energy within the vortex core have been observed experimentally, consistent with the calculated LDOS based on quasi-classical Green's function theory. Our findings stimulate a revision of the interpretation of previously reported non-split zero-energy bound states within vortex cores of different classes of superconducting materials.

**Figure 1.**

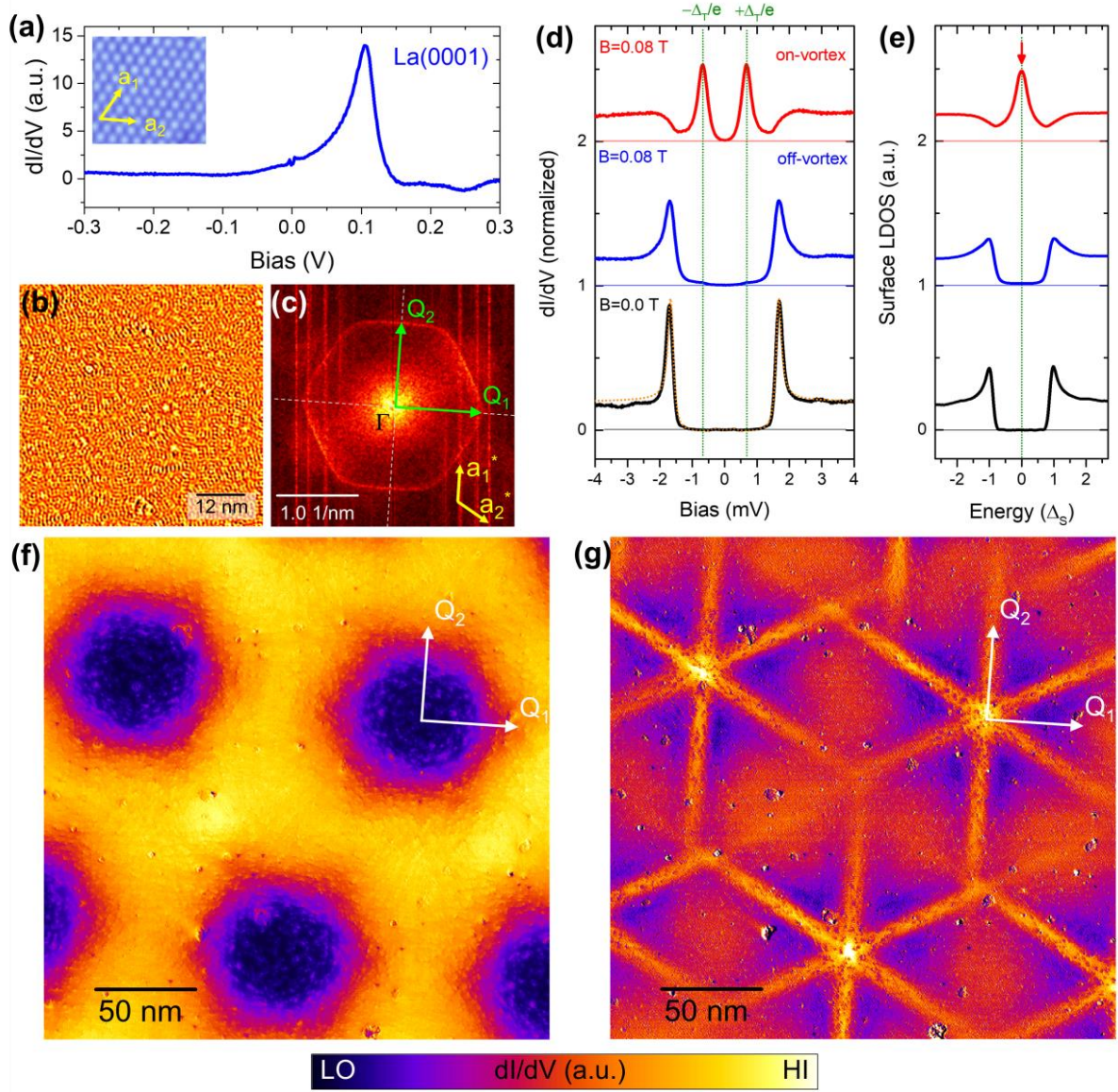

**Figure 1. Characterization of the surface electronic structure of La(0001) and visualization of superconducting vortices.**

**(a)** A typical differential tunneling conductance spectrum (dI/dV) measured on the La(0001) surface. The resonance peak at $V_S = +110$ mV is attributed to the edge of the surface band. The characteristic superconducting LDOS is visible at the Fermi energy. **(b)** A dI/dV map obtained at $V_S = +2$ mV showing quasiparticle interference (QPI) patterns. The image has been processed by Laplacian filtering for better visibility. **(c)** Fourier transform (FT) of the dI/dV map (raw data) at $V_S = +2$ mV showing a hexagonal scattering pattern, which can approximately be regarded as the Fermi contour of the 2D surface band. The arrows Q1 and Q2 denote the scattering vectors toward a vertex and a flat part of the hexagon, respectively. **(d)** dI/dV spectra obtained at the center of a superconducting vortex (on-vortex, red) and at the position between vortices (off-vortex, blue) by applying a magnetic field of 0.1 T perpendicular to the La surface. The black curve at the bottom shows a dI/dV spectrum without applying a magnetic field, while the red dotted curve represents a simulated spectrum for a superconductor-superconductor tunnel junction to obtain the superconducting gap edge of the tip ($\Delta_T$, vertical green lines). **(e)** The surface DOS plots obtained after numerical deconvolution of the tip DOS from the measured tunneling spectra shown in (d). The surface DOS at the vortex core shows a sharp zero-energy vortex bound state. Horizontal lines in (d) and (e) indicate the baselines of the zero conductance and DOS for each corresponding curve. **(f)** A differential tunneling conductance image at the superconducting coherence peak ($V_S = 1.6$ mV). Superconducting vortices are visible as depressions reflecting the suppression of superconductivity within the vortex cores. **(g)** A differential tunneling conductance image at $V_S = +\Delta_T/e = 0.65$ mV revealing a strongly anisotropic star-shaped distribution of the CBS.

**Figure 2.**

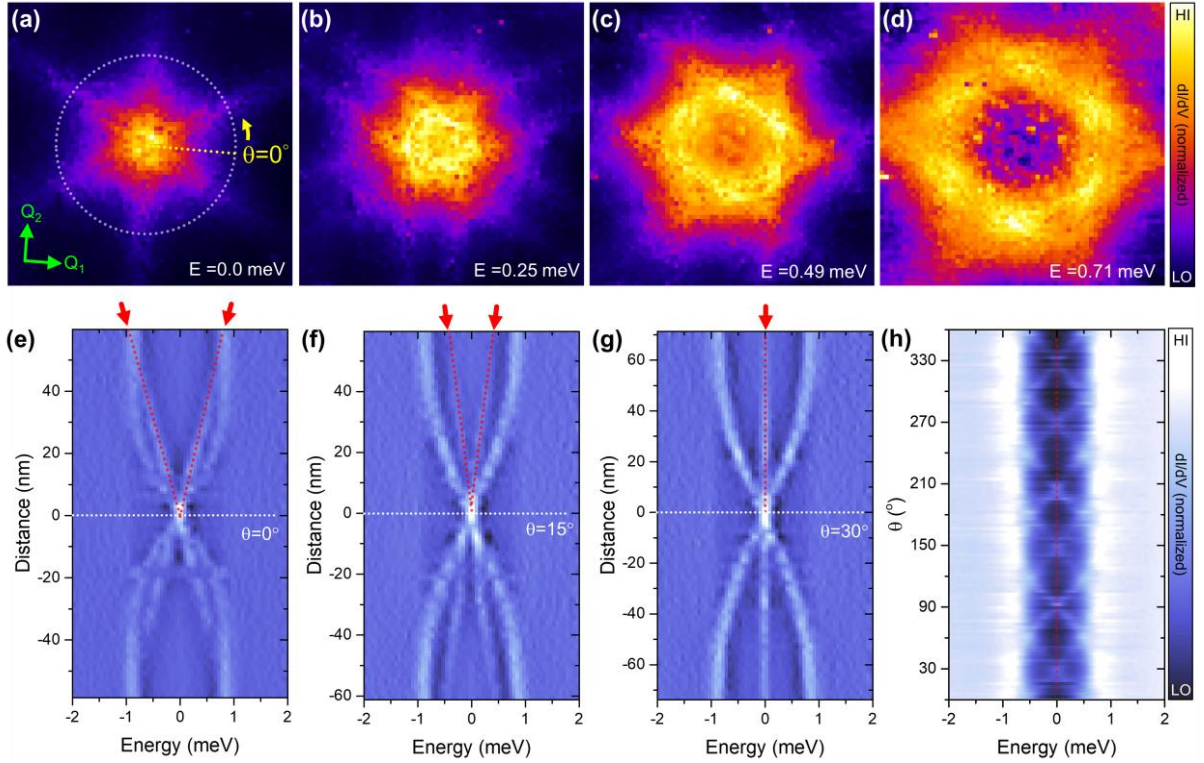

**Figure 2. Spatially- and energy-resolved distribution of the anisotropic vortex bound states on the La(0001) surface.**

**(a-d)** Spatially-resolved surface DOS maps for a single vortex at different energies: (a) E = 0 meV, (b) E = 0.25 meV, (c) E = 0.49 meV and (d) E = 0.71 meV. The green arrows, **Q₁** and **Q₂**, represent the directions of the scattering vectors in Fig. 1(c). Tunneling parameters: $I_T$ = 1.0 nA, $V_S$ = 2.5 mV, La-coated tip. Image size: 120×120 nm². **(e-g)** Spatial evolutions of the spectroscopic features along the radial lines for θ = 0° (e), θ = 15° (f) and θ = 30° (g) with respect to the **Q₁**-axis in Fig. 2(a). For better visibility of the evolution of the subgap features, the numerical derivatives of the dI/dV data are displayed [34]. Red dotted lines and arrows are guides to the eye for the inner and linear branches of the split VBS. Vertical and horizontal dotted lines are guidelines for zero-energy and the location of the vortex center, respectively. **(h)** Angle-dependent evolution of the spectroscopic features along the white dotted circle in Fig. 2A, 30 nm away from the vortex center. All tunneling spectra were obtained with a La-coated PtIr tip and have been normalized to the tunneling conductance at E = 2Δ_S. [See Methods].

**Figure 3.**

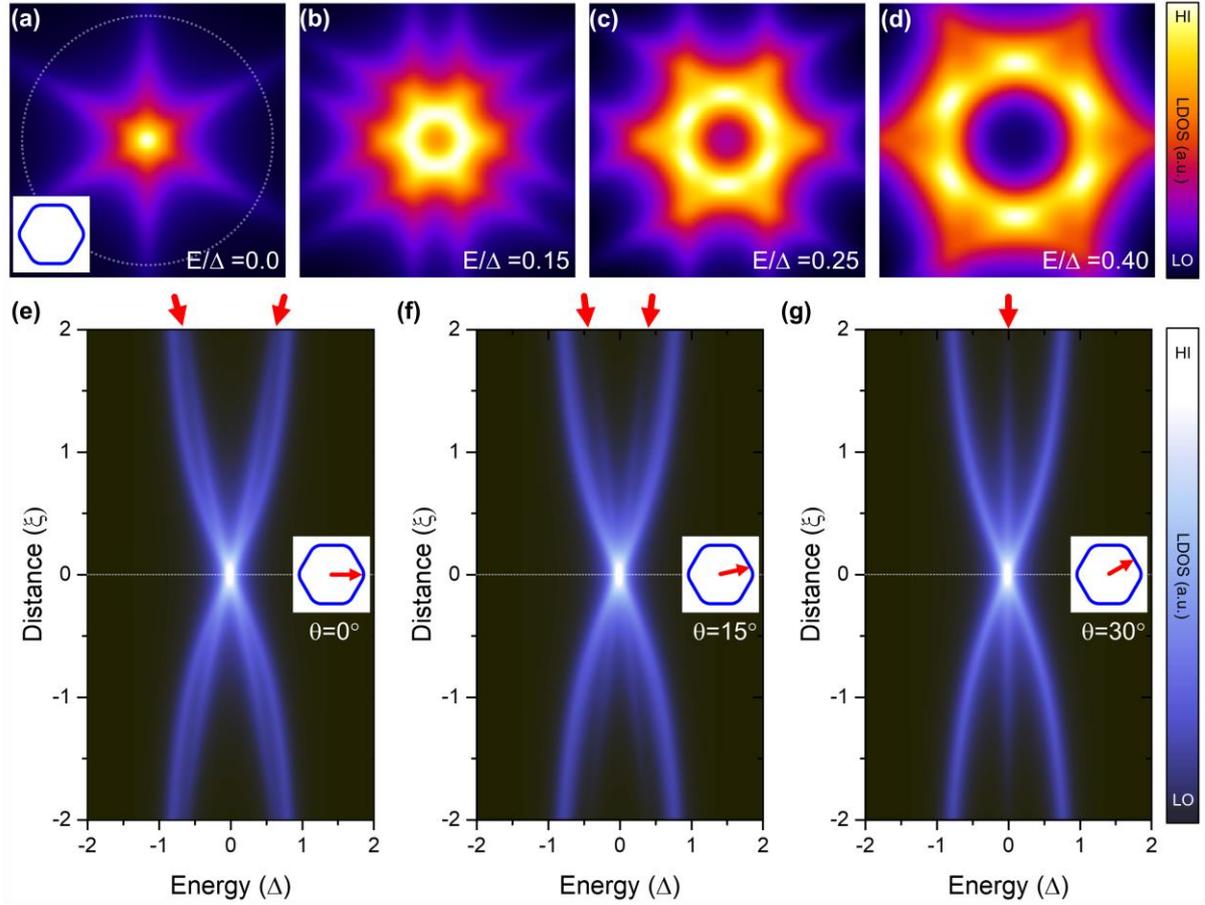

**Figure 3. Theoretically calculated quasiparticle DOS maps within a superconducting vortex based on the quasiclassical description.**

(**a-d**) Calculated LDOS maps (size: 1.1ξ × 1.1ξ) for a single isolated vortex as a function of energy inside the superconducting gap (Δ). The Fermi surface contour of the La(0001) surface extracted from the calculated band structure based on the DFT method is depicted in the inset of (a). The dotted line indicates a circle with radius 1.0ξ. Note that the superconducting order parameter model was considered to be isotropic for the calculations. (See the Methods Section for details). (**e-g**) Angle-resolved spatial evolutions of the vortex bound states along the radial directions (red arrows in the insets) across the vortex center for (e) θ = 0°, (f) θ = 15°, and (g) θ = 30° for the calculated Fermi surface contour.

*Acknowledgement*

We thank J. Wiebe, D. Wang and D. Morr for fruitful and useful discussions, as well as E. Simon, M. V. Valentyuk and A. I. Lichtenstein for discussions and support for first-principles calculations. This work was supported by the European Research Council via project No. 786020 (ERC Advanced Grant ADMIRE) at the University of Hamburg. L.R. is supported by the Alexander von Humboldt Foundation. This work was partially supported by the "Topological Materials Science" (No. 18H04228) JSPS-KAKENHI on Innovative Areas.

**APPENDIX: Experiments and theoretical methods**

**A. Preparation of the sample and tip.**

The rhenium single crystal was prepared by repeated cycles of $O_2$ annealing at 1400 K followed by flashing at 1800 K to obtain an atomically flat Re(0001) surface [37]. More than 50 nm thick lanthanum films were prepared in situ by electron beam evaporation of pure La pieces (99.9+%, MaTeck, Germany) in a molybdenum crucible at room temperature onto a clean Re(0001) surface, followed by annealing at 900 K for 10 minutes. The films were thicker than the coherence length of bulk La ($\xi$=36.3 nm) in order to suppress the inverse proximity effect from the Re. The surface cleanliness and the thickness of the La films were checked by STM after transferring the sample into the cryostat. A La-coated PtIr tip was used for STM/STS measurements to improve the energy resolution in the spectra at a given experimental temperature via the superconducting tip [34]. To fabricate the La-coated tip in situ, a mechanically polished PtIr tip was intentionally indented into the clean La film. The superconducting La-La junctions were confirmed by observing the Josephson tunneling current at V=0.0 mV as well as the spectral signature of multiple Andreev reflections.

**B. STM/STS measurements.**

All the experiments were performed in a low-temperature STM system (USM-1300S, Unisoku, Japan) operating at T=1.65 K under ultra-high vacuum conditions. Tunneling spectra were obtained by measuring the differential tunneling conductance (dI/dV) using a standard lock-in technique with a modulation voltage of 30 $\mu V_{rms}$ and a frequency of 1075 Hz with opened feedback loop. The bias voltage was applied to the sample and the tunneling current was measured through the tip using a commercially available controller (Nanonis, SPECS).

**C. Electronic structure calculations.**

The surface electronic structure of La(0001) was investigated using the screened Korringa–Kohn–Rostoker method [38]. The Fermi surface formed by the quasi-2D surface bands was determined via the calculation of the Bloch spectral function in the top La layer. Details of the calculations are reported in Ref. [27]. The symmetry-based indicators were calculated using the Check Topological Mat. program available on the Bilbao Crystallographic Database [32].

The topological classification of bulk La is available in existing online databases [32,33]. However, the nominal valence bands of La used in these databases are above the Fermi level in the vicinity of the Γ point, where the quasi-2D surface bands on the La(0001) surface are found. To deduce the topological character of these surface states, the nominal number of valence bands has to be reduced by four compared to the database value. Based on a recalculation of the electronic structure using the Vienna Ab-Initio Simulation Package [39–41], La was found to be an enforced semimetal also for lower nominal band filling, which does not indicate a topological nature for the observed surface states.

### D. Quasi-classical model calculations for the vortex bound states.

In the quasi-classical approximation, the Andreev scattered quasiparticles confined inside the vortex can be described by considering the quasiparticle trajectory which passes through the position characterized by the impact parameter [42]. The binding energy of the quasiparticles is a monotonically increasing and continuous function of the impact parameter, which is coupled to the angular momentum of the quasiparticle. At each spatial point $r$, the LDOS of the quasiparticles can be calculated by integrating over the number of quasiparticle trajectories passing through $r$ at a given impact parameter. Since the trajectory follows the direction of the Fermi velocity, which reflects the electronic band structure, this facilitates the correlation of the contribution of the Fermi surface with the spatial distribution of the quasiparticle LDOS around the vortex. The local density of states is calculated on the basis of the Kramer-Pesch approximation in the quasi-classical Eilenberger framework, which is appropriate in the low-energy regime [35]. The Fermi surface for these calculations was chosen to approximate the results of the ab initio calculations by interpolating between the perfectly circular and the ideal hexagonal Fermi surfaces using a deformation parameter. Details of this interpolation procedure are given in Ref. [27].